\documentclass[11pt]{article}

\usepackage{color,graphicx}
\usepackage{ifpdf}
\usepackage{amsmath}
\usepackage{amsthm}
\usepackage{amssymb}
\usepackage{cancel}

\usepackage
  [breaklinks,bookmarks,bookmarksnumbered,bookmarksopen,bookmarksopenlevel=2]
  {hyperref}
\hypersetup{pdftitle={Data Structures for Halfplane Proximity Queries and Incremental Voronoi Diagrams}}
\hypersetup{pdfauthor={Boris Aronov, Prosenjit Bose, Erik D. Demaine, Joachim Gudmundsson, John Iacono, Stefan Langerman, and Michiel Smid}}

\advance \topmargin by -\headheight
\advance \topmargin by -\headsep
\evensidemargin \oddsidemargin
{\makeatletter
 \gdef\xxxmark{%
   \expandafter\ifx\csname @mpargs\endcsname\relax 
     \expandafter\ifx\csname @captype\endcsname\relax 
       \marginpar{xxx}
     \else
       xxx 
     \fi
   \else
     xxx 
   \fi}
 \gdef\xxx{\@ifnextchar[\xxx@lab\xxx@nolab}
 \long\gdef\xxx@lab[#1]#2{{\bf [\xxxmark #2 ---{\sc #1}]}}
 \long\gdef\xxx@nolab#1{{\bf [\xxxmark #1]}}
 \long\gdef\xxx@lab[#1]#2{}\long\gdef\xxx@nolab#1{}%
}

\newtheorem{lemma}{Lemma} 
\newtheorem{problem}[lemma]{Problem} 
\newtheorem{opp}{Open Problem} 
\newtheorem{conj}[opp]{Conjecture} 
\newtheorem{theorem}[lemma]{Theorem} 
\newtheorem{corollary}[lemma]{Corollary} 
\newtheorem{definition}[lemma]{Definition} 
\newtheorem{transform}[lemma]{Transform}

\newcommand{\eps}{\varepsilon}               

\newcommand{\polylog}{\mathop{\rm polylog}}

\title{Data Structures for Halfplane Proximity Queries \\ and
  Incremental Voronoi Diagrams%
  \thanks{A preliminary version of this paper appeared in
    \emph{Proceedings of the 7th Latin American Symposium on
    Theoretical Informatics}, Valdivia, Chile, March 2006.}
}

\author{
Boris Aronov\thanks{Department of Computer Science and Engineering, Polytechnic School of Engineering, New York University, Brooklyn, NY, USA. Research supported in part by
NSF grant ITR-0081964 and by a grant from US-Israel Binational Science
Foundation.} \and 
Prosenjit Bose\thanks{School of Computer Science, Carleton University, Ottawa, ON, Canada. Research
supported in part by NSERC.} \and
Erik D. Demaine\thanks{Computer Science and Artificial Intelligence Laboratory,
MIT, Cambridge, MA, USA. Research supported in part by
NSF grants CCF-0430849 and OISE-0334653.} \and 
Joachim Gudmundsson\thanks{National ICT Australia, Sydney, Australia.} \and
John Iacono\thanks{Department of Computer Science and Engineering, Polytechnic School of Engineering, New York University, Brooklyn, NY, USA. Research
supported in part by NSF grants CCF-0430849, OISE-0334653,
CCF-1319648, CCF-1018370, and CNS-1229185, and a grant from
US--Israel Binational Science Foundation.} \and 
Stefan Langerman\thanks{Directeur de recherches du FNRS, 
D\'epartment d'Informatique, Universit\'e Libre de Bruxelles, Brussels, Belgium.} \and
Michiel Smid\footnotemark[3]
}

\date{}

\begin{document}

\maketitle

\begin{abstract}
  We consider preprocessing a set $S$ of $n$ points in convex position in the
  plane into a data structure supporting queries of the following form:
  given a point $q$ and a directed line~$\ell$ in the plane,
  report the point of $S$ that is farthest from
  (or, alternatively, nearest to) the point~$q$
  among all points to the left of line~$\ell$.
  We present two data structures for this problem.
  The first data structure uses $O(n^{1+\eps})$ space and preprocessing time,
  and answers queries in $O(2^{1/\eps}\log n)$ time, for any $0 < \eps < 1$.
  The second data structure uses $O(n \log^3 n)$ space and polynomial
  preprocessing time, and answers queries in $O(\log n)$ time.
  These are the first solutions to the problem with $O(\log n)$ query time
  and $o(n^2)$ space.

  The second data structure uses a new representation of nearest-
  and farthest-point Voronoi diagrams of points in convex position.
  This representation supports the insertion of new points in clockwise order
  using only $O(\log n)$ amortized pointer changes,
  in addition to $O(\log n)$-time point-location queries,
  even though every such update may make $\Theta(n)$ combinatorial changes
  to the Voronoi diagram.
  This data structure is the first demonstration that deterministically and
  incrementally constructed Voronoi diagrams can be maintained in
  $o(n)$ amortized pointer changes per operation while keeping $O(\log n)$-time
  point-location queries.
\end{abstract}

\section{Introduction}

Line simplification is an important problem in the area of digital
cartography \cite{Cromley-1991, Dent-1998, McMaster-Shea-1992}.
Given a polygonal chain~$P$,
the goal is to compute a simpler polygonal chain $Q$ that provides a good
approximation to $P$. Many variants of this problem arise depending
on how one defines \emph{simpler} and how one defines \emph{good
approximation}. Almost all of the known methods of approximation
compute distances between $P$ and $Q$. Therefore, preprocessing $P$
in order to quickly answer distance queries is a subproblem common
to most line simplification algorithms.

Of particular relevance to our work is a line simplification
algorithm proposed by Daescu et al.~\cite{dmsw-fpqgc-06}.  Given a
polygonal chain $P=(p_1, p_2, \ldots, p_n)$, they show how to compute
a subsequence $P'=(p_{i_1}, p_{i_2}, \ldots, p_{i_m})$,
with $i_1=1$ and $i_m=n$,
such that each segment $[p_{i_j}p_{i_{j+1}}]$ of $P'$ is a good
approximation of the subchain of $P$ from $p_{i_j}$ to $p_{i_{j+1}}$.
The amount of error is determined by the point of the subchain that
is farthest from the line segment $[p_{i_j}p_{i_{j+1}}]$. To compute
this approximation efficiently, the key subproblem they solve is
the following:

\begin{problem}[Halfplane Farthest-Point Queries]
  Preprocess $n$ points $p_1, p_2, \ldots, p_n$ in convex position in the plane
  into a data structure supporting the following query:
  given a point $q$ and a directed line $\ell$ in the plane, report
  the point $p_i$ farthest from $q$ among those to the left of line~$\ell$.
\end{problem}

Daescu et al.~\cite{dmsw-fpqgc-06} show that, with $O(n \log n)$
preprocessing time and space, these queries can be answered in
$O(\log^2 n)$ time.  On the other hand, a na\"{\i}ve approach achieves
$O(\log n)$ query time by using $O(n^3)$ preprocessing time and
$O(n^3)$ space.  A natural open question%
\footnote{Daescu et al.~\cite{dmsw-fpqgc-06} pose a closely related problem,
  whether $O(\log n)$ query time is possible with $O(n \log n)$ space and
  preprocessing time.}
is whether $O(\log n)$ query time can be obtained with a data structure
using subcubic and preferably subquadratic space.

In this paper, we solve this problem with two data structures.
The first, relatively simple data structure uses
$O(n^{1+\eps})$ preprocessing time and space, and answers queries
in $O(2^{1/\eps}\log n)$ time, for any $0 < \eps < 1$.
The second, more sophisticated data structure
uses $O(n \log^3 n)$ space and polynomial preprocessing time,
and answers queries in $O(\log n)$ time.
Both of our data structures apply equally well to halfplane farthest-point
queries, described above, as well as the opposite problem of halfplane
nearest-point queries.  Together we refer to these queries as
\emph{halfplane proximity queries}.

\paragraph{Dynamic Voronoi diagrams.}

An independent contribution of the second data structure is that it provides
a new efficient representation for maintaining the nearest- or
farthest-point Voronoi diagram of a dynamic set of points.
So far, point location in dynamic planar Voronoi diagrams has proved difficult
because the complexity of the changes to the Voronoi diagram or Delaunay
triangulation for an insertion can be linear at any one step.  The
randomized incremental construction avoids this worst-case behavior
through randomization.  However, for the deterministic insertion of
points, the linear worst-case behavior cannot be avoided, even if the
points being incrementally added are in convex position, and are added
in order (say, clockwise).  For this specific case, we give a
representation of a (nearest- or farthest-point) Voronoi diagram that
supports $O(\log n)$-time point location in the diagram while requiring only
$O(\log n)$ amortized pointer changes in the structure for each update.
So as not to oversell this result, we note that we do not have an efficient
method of determining which pointers to change (it takes $\Theta(n)$ time
per change), so the significance of this representation is that it serves
as a proof of the existence of an encoding of Voronoi diagrams that can be
modified with few changes to the encoding while still supporting
point-location queries.

Since the conference version of this paper first appeared, there have been
two significant follow-up works.
First, Allen et al.~\cite{superflarb_DCG} showed how to compute the necessary
$K$ topological changes in the Voronoi diagram in $O(K \polylog n)$ time.
In particular, this result implies an explicit preprocessing algorithm for
building our second data structure that runs in $O(n \polylog n)$ time.
Second, Pettie~\cite{pettie} gave a simpler proof of our $O(\log n)$ upper
bound on the number of topological changes, and proved a matching lower bound
when following the same combinatorial approach and not exploiting any further
geometry.

Currently, the best incremental data structure supporting nearest-neighbor
queries (one interpretation of ``dynamic Voronoi diagrams'')
supports queries and insertions in $O(\log^2 n / \log \log n)$.
This result uses techniques for decomposable search problems
described by Overmars~\cite{o-ddds-83}; see \cite{ct-dacg-92}.
More recently, Chan \cite{c-dds3c-10} developed a randomized data structure
supporting nearest-neighbor queries in $O(\log^2 n)$ time,
insertions in $O(\log^3 n)$ expected amortized time,
and deletions in $O(\log^6 n)$ expected amortized time.
By contrast, our data structure for points in convex position added in
clockwise order achieves $O(\log n)$ query time, $O(\polylog n)$ insert time,
$O(n \log^3 n)$ space, and $O(n \polylog n)$ preprocessing time.

\section{A Simple Data Structure}

\def\D{\mathcal{D}}


When referring to some or all of $n$ points in convex position and
clockwise order $p_1, p_2, \dots, p_n$, the indices are to be understood
modulo~$n$, and $p_{[i,j]}$ refers to the contiguous sequence of points
going clockwise from $p_i$ to $p_j$, wrapping around $p_n$ and $p_1$
if $j<i$.

In this section, we prove the following theorem:

\begin{theorem} \label{simple theorem}
  There is a data structure for halfplane proximity queries on a static
  set of $n$ points in convex position that achieves $O(2^{1/\eps} \log n)$
  query time using $O(n^{1+\eps})$ space and preprocessing,
  for any $0 < \eps < 1$.
\end{theorem}

Our proof is based on starting from the na\"{\i}ve $O(n^3)$-space
data structure mentioned in the introduction, and then repeatedly applying
a space-reducing transformation.
We assume that either all queries are halfplane farthest-point queries
or all queries are halfplane nearest-point queries; otherwise, we can
simply build two data structures, one for each type of query.

Both the starting data structure and the reduction use Voronoi diagrams as
their basic primitive.  More precisely, we use the farthest-site Voronoi
diagram for the case of halfplane farthest-point queries, and the nearest-site
Voronoi diagram for the case of halfplane nearest-point queries.
When the points are in convex position and given in clockwise order,
Aggarwal et al.~\cite{agss-ltacv-89} showed that either Voronoi diagram can be
constructed in linear time.  Answering point-location queries in either
Voronoi diagram of points in convex position can be done in $O(\log n)$ time
using $O(n)$ preprocessing and space~\cite{egs-oplms-86}.

\begin{lemma} \label{obs:basicAlg}
  There is a static data structure for halfplane proximity queries on a static
  set of $n$ points in convex position, called \emph{Okey},
  that achieves $O(\log n)$ query time using $O(n^3)$ space and
  preprocessing.
\end{lemma}

\begin{proof}
  Let $p_1, p_2, \dots, p_n$ denote the $n$ points
  in convex position in clockwise order.
  The Okey data structure consists of one Voronoi diagram $V(i,j)$ for every
  contiguous subsequence $p_{[i,j]}$ of points.
  (This exact data structure was suggested in the conclusion of
   \cite{dmsw-fpqgc-06}, but without details or analysis.)
  The space and preprocessing is thus $O(n^3)$.

  To answer a halfplane proximity query for a point $q$
  and a directed line~$\ell$, we first find the subsequence of points
  on the left of line~$\ell$.
  In $O(\log n)$ time, we can determine whether $\ell$ intersects the
  convex hull, and if so, find the two edges $(p_i, p_{i+1})$ and
  $(p_j, p_{j+1})$ of the convex hull that are intersected by the query line
  \cite[Section~7.9.1]{O'Rourke-1998-C}.
  Then, depending on the orientation of the line~$\ell$ (i.e., which edge is
  struck first), we can decide between the two possible intervals:
  $p_{[i+1,j]}$ or $p_{[j+1, i]}$.
  Then we locate $q$ in the appropriate Voronoi diagram,
  either $V(i+1,j)$ or $V(j+1,i)$ (or $V(1,n)$ if $\ell$ does not
  intersect the convex hull), and return the site $p_k$ that
  generated the corresponding Voronoi region.
  The total query time is $O(\log n)$.
\end{proof}

\begin{transform}%
  \hspace*{-0.75em}
  \footnote{We use the term ``Transform'' to denote a type of theorem
    that represents a data structure transformation.}
  \label{shrink transform}
  Given any static data structure $\D$ for halfplane proximity queries
  on a static set of $n$ points in convex position that achieves $Q(n)$ query
  time using $M(n)$ space and preprocessing, and for any parameter $m \leq n$,
  there is a static data structure for halfplane proximity queries on a static
  set of $n$ points in convex position, called $\D$-\emph{Dokey}, that
  achieves $2 Q(n) + O(\log n)$ query time using
  $\lceil n/m \rceil \, M(m) + O(n^2/m)$ space and preprocessing.
\end{transform}

\begin{proof}
  Let $p_1, p_2, \ldots, p_n$ be the $n$ points
  in convex position in clockwise order.
  We define the $\lceil n/m \rceil$ \emph{breakpoints} to be the points $p_k$
  with $k \equiv 1 \pmod{m}$, i.e., the points
  $p_{i m + 1}$ for $i \in \{0, 1, \dots, \lceil n/m \rceil-1\}$.
  The data structure consists of two substructures:
  \begin{description}
  \item[\boldmath $\D$ Substructure:]
    We construct an instance of the data structure $\D$ on the half-open
    interval of points between every consecutive pair of breakpoints.
    More precisely, for each $i \in \{0, 1, \dots, \lceil n/m \rceil-1\}$,
    we construct an instance of $\D$ on the points
    $p_{[i m+1, \min\{n,(i+1)m\}]}$.
    These structures require $\lceil n/m \rceil (M(m) + O(1))$
    space and preprocessing.
  \item[Voronoi Substructure:]
    For each breakpoint $p_k$, we construct Voronoi diagrams on
    all intervals of points of length an exact power of two with
    one endpoint at~$p_k$.
    More precisely, for each $i \in \{0, 1, \dots, \lceil n/m \rceil-1\}$,
    and for each $j \in \{0, 1, \dots, \lfloor \log n \rfloor\}$,
    we construct two Voronoi diagrams, one on the points $p_{[i m+1,i m+2^j]}$,
    and one on the points
    $p_{[i m+1,i m+2-2^j]}$. 
    The space and preprocessing requirements for these Voronoi diagrams are
    $$O\left( {n \over m} \cdot \sum_{j=0}^{\lfloor \log n \rfloor} 2^j \right)
     = O\left( {n^2 \over m} \right).$$
  \end{description}
  Overall, the space and preprocessing required for $\D$-Dokey is
  $\lceil n/m \rceil \, M(m) + O(n^2/m)$ as claimed.

  It remains to show how we can use $\D$-Dokey to answer halfplane proximity
  queries in $2 Q(n) + O(\log n)$ time.
  Suppose that we are given a point $q$ and a directed line~$\ell$.
  As described in the proof of Lemma~\ref{obs:basicAlg}, in $O(\log n)$ time,
  we can find the interval $p_i, p_{i+1}, \ldots, p_j$ of points
  to the left of line~$\ell$.
  If this interval contains no breakpoints, then it is contained in the
  interval of a $\D$ substructure, so we can answer the query in $Q(n)$ time
  by passing it to the $\D$ substructure. 
  Otherwise, let $p_{i'}$ and $p_{j'}$ be the first and the last breakpoints
  in the interval, respectively.
  We ask the $\D$ substructure immediately preceding $p_{i'}$
  (representing the interval $p_{[i'-m, i'-1]}$
  if $i' > 0$, and the interval
  $p_{[(\lceil n/m \rceil-1) m+1, n]}$
  if $i' = 0$)
  and the $\D$ substructure immediately succeeding $p_{j'}$
  (representing the interval $p_{[j', \min\{n,j'+m-1\}]}$)
  the same halfplane proximity query.
  These queries cover the ranges $p_{[i, i'-1]}$
  and $p_{[j', j]}$.
  To cover the remaining range $p_{[i', j']}$ between the
  two breakpoints, we use the property that any interval can be covered
  (with overlap) by two intervals of length an exact power of two.
  Namely, let $k = 2^{\lfloor \lg (j'-i') \rfloor}$,
  where the difference $j'-i'$ accounts for wraparound modulo~$n$.
  We query $q$ in the Voronoi diagram on the interval
  $p_{[i', i'+k]}$
  and in the one on the interval
  $p_{[j'-k, j']}$.
  Together, the four queries cover (with overlap) the desired interval
  $p_{[i, j]}$.
  Among the four results from the four queries,
  we return the best (either farthest or nearest) relative to point~$q$.
\end{proof}

By starting with the data structure Okey of Lemma~\ref{obs:basicAlg},
and repeatedly applying the Dokey transformation of
Transformation~\ref{shrink transform},
we obtain the structure Okey-Dokey-Dokey-Dokey-\dots,
or Okey-Dokey$^k$, which leads to the following:

\begin{corollary} \label{simple corollary}
  For every integer $k \geq 1$, \emph{Okey-Dokey}$^{k-1}$ is a data structure for
  halfplane proximity queries on a static set of $n$ points in convex
  position that achieves $O(2^k \log n)$ query time using
  $O(k \, n^{(2 k+1)/(2 k-1)}) = O(k \, n^{1+1/(k-1/2)})$ space and preprocessing.
\end{corollary}

\begin{proof}
  The proof is by induction on~$k$.
  In the base case $k=1$, we can use the Okey data structure
  from Lemma~\ref{obs:basicAlg} because $(2 k+1)/(2 k-1) = 3$.
  For $k>1$, assume by induction that we have a data structure for $k-1$
  that achieves query time at most $c (2^k-1) \log n$
  using space and preprocessing at most $c (k-1) n^{(2 k-1)/(2 k-3)}$.
  Assume that the constant $c$ is at least twice as large as the constants
  implicit in the $O$ notation in Transform~\ref{shrink transform}.
  We apply the Dokey transformation from Transform~\ref{shrink transform}
  to this data structure, substituting $m = n^{(2 k-3)/(2 k-1)}$.
  Thus, $n/m = n^{2/(2 k-1)}$ and $n^2/m = n^{(2 k+1)/(2 k-1)}$.
  The resulting query time is at most
  $2 c  (2^k-1) \log n + (c/2) \log n
  \leq c  (2^{k+1}-1) \log n$, as desired.
  The resulting space and preprocessing time is at most
  $(n/m+1)  c (k-1) m^{(2 k-1)/(2 k-3)} + (c/2)  n^2/m
  = c  (n^{2/(2 k-1)}+1) (k-1) n + (c/2)  n^{(2 k+1)/(2 k-1)}
  = c (k-1) n^{(2 k+1)/(2 k-1)} + c (k-1) n + (c/2)  n^{(2 k+1)/(2 k-1)}
  \leq c \, k \, n^{(2 k+1)/(2 k-1)}$ for sufficiently large~$n$, as desired.
\end{proof}

The space and preprocessing time of Okey-Dokey$^{k-1}$ according to
Corollary~\ref{simple corollary} can be written as $n^{1 + 2/(2 k-1)}$.
For any given $\eps>0$, we choose $k = \lceil 1/2 + 1/\eps \rceil$.
Then the space and preprocessing time are $O(n^{1+\eps})$
and the query time is $O(2^{1/\eps} \log n)$,
proving Theorem~\ref{simple theorem}.

\section{Grappa Trees}

\begin{figure}
  \centering
  \scalebox{0.6}{\ifpdf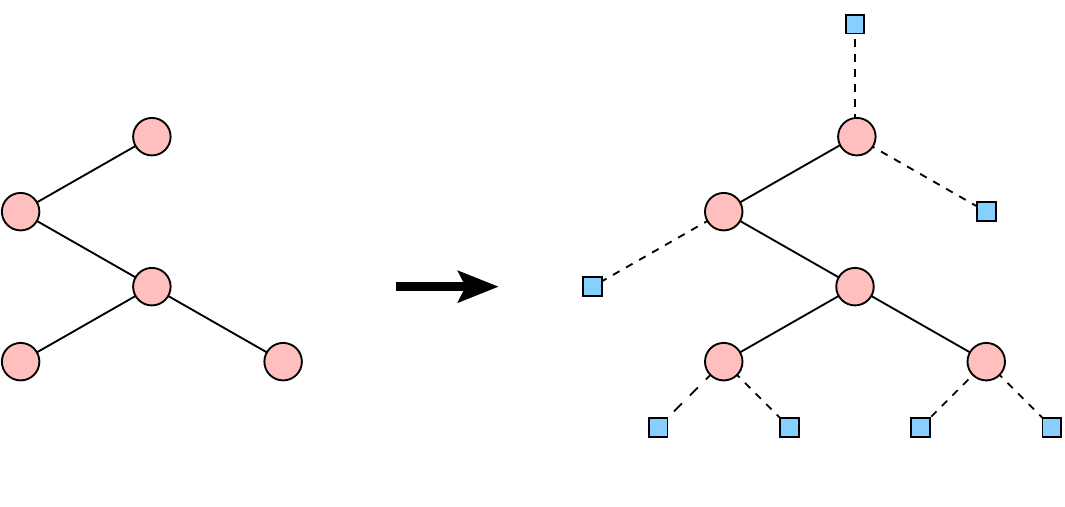\else\input{expanded.pstex_t}\fi}
  \caption{Expanding a rooted binary tree with external nodes and a superroot
    (squares).  Internal nodes are circles.}
  \label{fig:expanded}
\end{figure}

Our faster data structure for halfplane proximity queries requires
the manipulation of binary trees with topology
determined by a Voronoi diagram.
To support efficient manipulation of such trees, we introduce a data structure
called \emph{grappa trees}.  This data structure is a modification
of Sleator and Tarjan's link-cut trees \cite{Sleator-Tarjan-1983}
that supports some unusual additional operations.
For convenience, we \emph{expand} a given rooted binary tree
by adding an \emph{external} vertex in place of each absent left/right child
and adding a \emph{superroot} vertex above the root;
see Figure~\ref{fig:expanded}.
Thus every \emph{internal} (original) vertex is incident to exactly three edges,
and the external vertices and superroot are the leaves.
(A slight aberration: although the root is the child of the superroot,
 it is neither a left nor right child.)

\begin{definition}
  Grappa trees solve the following data-structural problem:
  maintain a forest of expanded rooted binary trees
  with specified topology, and a ``left mark'' and ``right mark''
  on each edge, subject to
  \begin{description}
  \item [\rm $T$ = Make-Tree$(v)$:] Create a new tree $T$ with a single
    internal vertex~$v$ (not previously in another tree).
    Implicitly this operation also creates two external vertices
    and a superroot for~$v$, and three edges with null labels.
  \item [\rm $T$ = Link$(v,w)$:]
    Given an external vertex $v$ in one tree $T_v$ and the superroot $w$ of
    a different tree~$T_w$, connect the parent $v'$ of $v$ via an edge to
    the root child $w'$ of $w$, deleting the extra nodes $v$ and~$w$, and
    merging $T_v$ and $T_w$ into a new tree~$T$.
    The new edge $(v',w')$ is assigned the left and right marks of $(w,w')$.%
    \footnote{This convention is arbitrary, but it allows first setting the
      marks of $(w,w')$ via Left-Mark and Right-Mark to effectively set the
      marks of $(v',w')$.}
  \item [\rm $(T_1,T_2)$ = Cut$(e)$:] Delete the existing edge $e = (v,w)$
    in tree~$T$, splitting $T$ into two trees $T_1$ and $T_2$
    containing $v$ and $w$, respectively.
    If say $T_1$ contains the superroot of~$T$,
    then $v$ gains a new external child (replacing the connection to~$w$),
    and $w$ becomes a root and gains a new superroot parent
    (replacing the connection to~$v$).
    The new edges incident to $v$ and $w$ acquire the same left and right
    labels as the original edge~$e$.
  \item [\rm Evert$(v)$:] Make external node $v$ the superroot of its tree,
    reversing the orientation (which endpoint is closer to the superroot)
    of every edge along the superroot-to-$v$ path.
    The left/rightness of each child/edge is uniquely determined by preserving
    the cyclic order of edges around each vertex.
  \item [\rm Left-Mark$(T,v,m_\ell)$:]
    Set the left mark of every edge on the superroot-to-$v$ path in $T$ to
    the new mark $m_\ell$, overwriting the previous left marks of these edges.
  \item [\rm Right-Mark$(T,v,m_r)$:]
    Set the right mark of every edge on the superroot-to-$v$ path in $T$ to
    the new mark $m_r$, overwriting the previous right marks of these edges.
  \item [\rm $(e,m^*_\ell,m^*_r)$ = Oracle-Search$(T,O_e)$:]
    Search for the edge $e$ in tree~$T$.
    The data structure can find $e$ only via \emph{oracle queries}:
    given two incident edges $f$ and $f'$ in~$T$, the provided oracle
    $O_e(f,f',m_\ell,m_r,m'_\ell,m'_r)$ determines in constant time
    which ``side'' of $f$ contains $e$, i.e., whether $e$ is in the component
    of $T - f$ that contains~$f'$, or in the rest of the tree
    (which includes $f$ itself).%
    %
    \footnote{Given the number of arguments, it is tempting to refer to the
      oracle as $O_A(B,D,G,I,L,S)$, but we will resist that temptation.}
    %
    The data structure provides the oracle with the left mark $m_\ell$
    and the right mark $m_r$ of edge $f$, as well as the left mark $m'_\ell$
    and the right mark $m'_r$ of edge $f'$, and at the end, it returns
    the left mark $m^*_\ell$ and the right mark $m^*_r$ of the found edge~$e$.
  \end{description}
\end{definition}


\begin{theorem} \label{grappa theorem}
  There exists an $O(n)$-space constant-in-degree pointer-machine data
  structure that maintains a forest of grappa trees and supports each operation
  in $O(\log n)$ worst-case time per operation, where $n$ is the total size
  of the trees affected by the operation.  (In fact, for the time bound,
  $n$~can be just the total size of the trees involved in the operation.)
\end{theorem}

\begin{proof}
  Our grappa-tree data structure is based on the worst-case version of the
  link-cut tree data structure of Sleator and Tarjan
  \cite[Section~5]{Sleator-Tarjan-1983}.
  This data structure maintains a forest of specified-topology trees
  subject to Make-Tree, Link, Cut, and several other operations,
  each in $O(\log n)$ worst-case time per operation, and using $O(n)$ space.
  The data structure represents each tree in the forest
  by decomposing it into a set of maximal vertex-disjoint downward paths,
  connected by tree edges called \emph{nonpath edges}.
  Each path is in turn represented by a biased binary tree
  whose leaf nodes represent the vertices of the path,
  and whose nonleaf nodes represent the edges of the path,
  ordered in the biased tree according to the depth along the path.
  Thus, vertices of larger height in the path correspond to leaf nodes
  farther left in the biased tree.
  For each leaf node $v$ of a biased tree representing an internal vertex $u$
  in~$T$,
  $u$~has a unique nonpath child edge (because paths are maximal and $T$ is
  an expanded rooted binary tree), which we can associate with~$v$.
  The link-cut tree structure for an expanded rooted binary tree $T$
  can therefore be seen as a rooted tree $R$, the \emph{representation tree},
  in which every node corresponds to an edge of~$T$.
  A node that is a nonleaf of its biased tree
  represents a path edge and has exactly two children, while
  a node that is a leaf of its biased tree
  represents a nonpath edge and has at most one child.
  Thus we call these two types of nodes \emph{path nodes} and
  \emph{nonpath nodes}, respectively.
  For the subtree $R_v$ of $R$ rooted at a node~$v$, the nodes of $R_v$
  correspond to edges in $T$ that form a connected subtree~$T_v$ (namely,
  an interval of the path containing the edge of $T$ represented by~$v$,
  plus the nonpath children edges and their rooted subtrees in~$T$).
  By a suitable choice of paths and biasing,
  as described in \cite{Sleator-Tarjan-1983},
  $R$~has height $O(\log n)$.

  We augment the representation tree $R$ to enable marking as follows.
  Because our tree $T$ has bounded degree (an assumption not made
  in~\cite{Sleator-Tarjan-1983}), we can also explicitly store $T$
  (the parent, left child, and right child of each vertex) and cross-link
  corresponding nodes/vertices and corresponding edges in the two structures.
  To each edge of $T$ we add a left-mark field and a right-mark field.
  These fields contain the last explicitly stored marks for the edge,
  and for nonpath edges, they are accurate,
  while for path edges, the mark fields may become out-of-date.
  To each path node in~$R$, we also add a
  left-mark field and a right-mark field, which may be blank.
  When nonblank, each field represents bulk markings that should be
  (but have not yet been) applied to the descendant path nodes within
  the same biased tree.
  Thus, the actual left mark of an edge $e$ on a path in $T$ is implicitly
  the first nonblank left-mark field of a node along the path from the root of
  the biased tree representing the path containing $e$, if there is such a
  nonblank field, or else the left-mark field of the edge
  $e$ itself; and symmetrically for right marks.

  We can maintain this augmentation as the representation tree $R$ changes.
  Because the definition of the augmented values is relative to individual
  biased trees, we care only about modifications to biased trees themselves,
  not about the modifications to the edges between different biased trees
  that form the entire representation tree~$R$.
  The link-cut data structure modifies biased trees according to rotations,
  splits, and concatenations.
  We can modify the implementation of all of these operations to propagate
  the mark fields, at the cost of an extra constant factor,
  in such a way that preserves the implicit marks of all edges in~$T$.
  The idea is to push down node marks judiciously:
  whenever any operation visits a path node $v$ of $R$
  with a nonblank mark field, copy that value to the corresponding mark field
  of the edge of $T$ represented by~$v$, as well as to the mark field
  of any child of $v$ that is a path node in~$R$
  (overwriting any previous value),
  and finally blank out the field in the node $v$ itself.
  Because operations on link-cut trees always start at the root of~$R$
  and traverse along paths down from there, any nodes involved in the operation
  will have already cleared their mark fields before they actually get used,
  so the marks on the corresponding edges in $T$ will be up-to-date.

  To implement Left-Mark$(T,u,m)$ or Right-Mark$(T,u,m)$, we visit all biased
  trees that represent paths containing edges along the superroot-to-$u$ path
  in~$T$.
  We start with the bottommost edge from $u$ to its parent in~$T$,
  and its corresponding node $v$ in~$R$.
  Then we walk up $R$ from~$v$.
  Whenever we walk from a right child $v$ to its parent $w$ that is a path
  node in~$R$, we set the appropriate (left- or right-) mark field of
  $w$'s left child in $R$ to~$m$
  (because all descendant leaf nodes in the biased tree are left of $w$
   so correspond to edges of the $T$ path above~$v$);
  we also set the appropriate mark field of the edge of $T$ represented by $w$
  to~$m$.
  Whenever we walk through a nonpath node $w$ of~$R$,
  we set the appropriate mark field of the edge of $T$ represented by~$w$.
  Because $R$ has height $O(\log n)$, the entire length of the walk
  and thus the total number of markings is $O(\log n)$.

  Given a query oracle $O_e$ and a tree~$T$, we can perform Oracle-Search
  by a tree walk in $R$ starting at the root.  Upon visiting a path node $v$
  of $R$ representing a path edge $f=(u,w)$ of $T$, we find the nonpath child
  edges $f'$ of~$u$ and $f''$ of~$w$ (both incident to~$f$).
  Because $f$ was just visited, its mark fields $m_\ell$ and $m_r$ will be
  up-to-date, and because $f'$ and $f''$ are nonpath edges of $T$,
  their mark fields $m'_\ell,m''_\ell$ and $m'_r,m''_r$ are accurate.
  Thus we can make two calls to the oracle---$O_e(f,f',m_\ell,m_r,m'_\ell,m'_r)$
  and $O_e(f,f'',m_\ell,m_r,m''_\ell,m''_r)$---to determine whether
  $f$ is the edge $e$ we are looking for, or else which of the two child
  subtrees of $v$ in $R$ contains the node representing~$e$.
  In the special cases when $u$ or $w$ is an external node or the superroot
  of~$T$, $f'$~or $f''$ does not exist, and we only need to perform one of the
  tests: if the oracle points to the side containing~$f$, then $f=e$.
  Upon visiting a nonpath node $v$ of~$R$ representing a nonpath edge $(u,w)$
  in~$T$, where $w$ is the parent of~$u$, we find the nonpath child edge
  $f'$ of~$u$, and call $O_e(f,f',m_\ell,m_r,m'_\ell,m'_r)$, to determine
  whether $f=e$ or $e$ is in the subtree $R_w$.
  In the special case when $u$ is an external node of~$T$,
  $f'$ does not exist, but then we know that $f=e$ without any oracle calls.
  Because $R$ has height $O(\log n)$,
  Oracle-Search queries run in $O(\log n)$ worst-case time.
\end{proof}

\section{Rightification of a Tree: Flarbs}

The specified-topology binary tree maintained by our faster data
structure for halfplane proximity queries changes in a particular way
as we add sites to a Voronoi diagram.
We delay the specific connection for now,
and instead define the way in which the tree changes:
a tree restructuring operation called a ``flarb''.
Then we bound the work required to implement a sequence of $n$ flarbs
by showing that the total number of pointers changes
(i.e., the total number of parent/left-child and parent/right-child
relationships that change) is $O(n \log n)$.
Thus, for the remainder of this section, we use the term \emph{cost}
to refer to (a constant factor times) the number of pointer changes
required to implement a tree-restructuring operation,
not the actual running time of the implementation.
This bound on cost will enable us to implement a sequence of $n$ flarbs
via $O(n \log n)$ link and cut operations, for a total of
$O(n \log^2 n)$ time.

The flarb operation is parameterized by an ``anchored subtree''
which it transforms into a ``rightmost path''.
An \emph{anchored subtree} $S$ of a nonempty rooted binary tree $T$ is any
connected subgraph $S$ of $T$ that includes the root of~$T$;
in the special case of an empty tree $T$,
we define an anchored subtree $S$ of $T$ to be the empty subgraph.
A \emph{right-leaning path} in a rooted binary tree $T$ is a path
monotonically descending through the tree levels,
always proceeding from a node to its right child.
A \emph{rightmost path} in $T$ is a right-leaning path
that starts at the root of~$T$.

\begin{figure}
  \centering
  \scalebox{0.45}{\ifpdf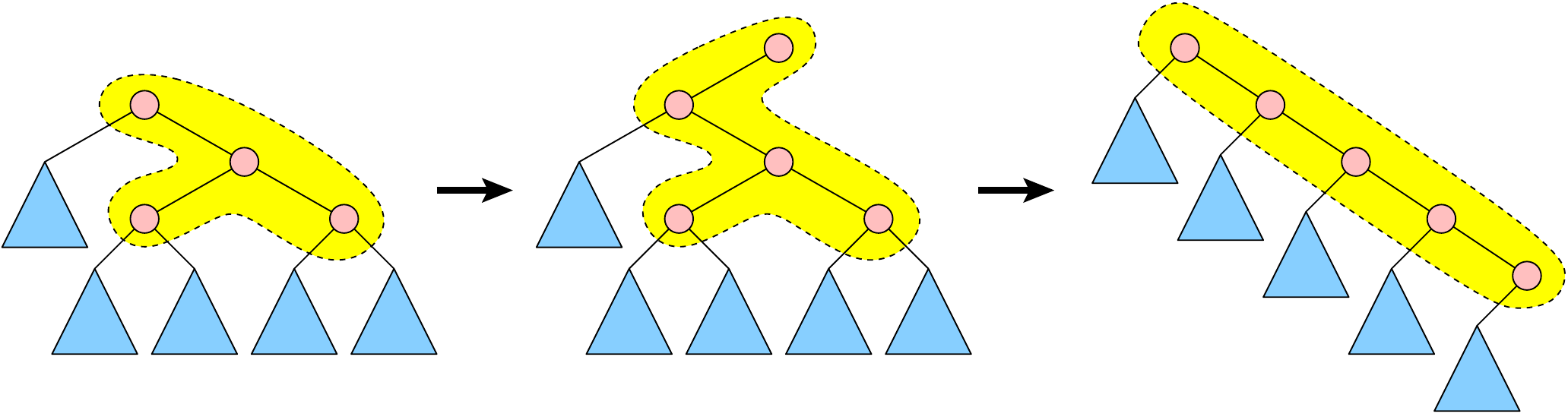\else\input{ex.pstex_t}\fi}
  \caption{An example of a flarb.
           The anchored subtree is highlighted.}
  \label{fig:ex}
\end{figure}

The \emph{flarb} operation%
\footnote{%
  Note that this notion of flarb is different from
  that of~\cite{flarb}. \label{flarb footnote}}
of an anchored subtree $S$ of a rooted binary tree $T$
is a transformation of $T$ defined as follows;
refer to Figure~\ref{fig:ex}.
First, we create a new root node $r$ with no right child and
whose left child subtree is the previous instance of~$T$;
call the resulting rooted binary tree~$T'$.
We extend the anchored subtree $S$ of $T$ to an anchored subtree
$S'$ of $T'$ by adding $r$ to~$S$.
Now we re-arrange $S'$ into a rightmost path on the same set of nodes,
while maintaining the in-order traversal (binary search tree order)
of all nodes.
The resulting rooted binary tree $T''$ is the result of flarbing $S$ in~$T$.

Now we consider a sequence of flarb operations $f_1, f_2, \dots$,
where $f_1$ applies to an empty tree $T_0$, and each flarb operation
$f_i$ transforms each successive tree $T_{i-1}$ into~$T_i$.
Note that each flarb can choose a different anchored subtree~$S_i$.
Although the size of the anchored subtrees may be very large,
we show that the number of actual pointer changes is small:

\begin{theorem} \label{flarb theorem}
  A sequence of $n$ flarb operations, starting from an empty tree,
  can be implemented at a cost of $O(\log n)$ amortized pointer changes
  per flarb.
\end{theorem}

\begin{proof}
  We use the potential method of amortized analysis,
  with a potential function inspired by the analysis of splay trees
  \cite{Sleator-Tarjan-1985-splay}.
  For any node $x$ in a tree~$T$,
  let $w(x)$ be the \emph{expanded weight} of the subtree rooted at~$x$,
  which is the number of nodes in the subtree plus the number
  of null pointers in the tree.  In other words, as in expanded trees,
  we add external nodes in place of each null pointer in~$T$,
  but here just for the purpose of computing subtree size.
  Define $\varphi(x)= \lg \frac{w(\textit{left}(x))}{w(\textit{right}(x))}$.
  Clearly $|\varphi(x)| \leq \lg (2n-1)$,
  because the smallest possible subtree contains no
  real nodes and one external node, and the largest possible subtree
  contains $n-1$ real nodes and $n$ external nodes.
  The potential of a tree $T$ with $n$ nodes is
  $\Phi(T) = \sum_x \varphi(x)$, with the sum
  taken over the (real) nodes $x$ in~$T$.
  Therefore, $|\Phi(T)| = O(n \log n)$ for any tree~$T$.

  For the purposes of the analysis, we use the following heavy-path
  decomposition of the tree.
  The \emph{heavy path} from a node continues recursively 
  to its child with the larger subtree (breaking ties arbitrarily),
  and the \emph{heavy-path decomposition} is the natural decomposition
  of the tree into maximal heavy paths.
  Edges on heavy paths are called \emph{heavy edges},
  while all other edges (connecting two heavy paths) are called
  \emph{light edges}.

\paragraph{Outline.}

To analyze a flarb in a rooted binary tree $T$, we decompose the
transformation into a sequence of several steps,
and analyze each step separately.

First, the addition of the new root node $r$ can be performed by changing a
constant number of pointers in the tree.
Because $\varphi(r) = \lg (2 n - 1)$, the amortized cost of this
operation is trivially $O(\log n)$.
Thus, in the remainder of the proof, we focus on the actual restructuring
of the resulting anchored subtree $S'$ into a rightmost path,
a process we call \emph{rightification}.

At all times during rightification, the nodes constituting the
original anchored subtree $S'$ continue to form an anchored subtree
of the current rooted binary tree, and for simplicity of notation
we continue to denote the current such anchored subtree as~$S'$.

To implement rightification, we first execute several simplifying steps
of two types, called ``zig'' and ``zag'',%
\footnote{Unlike most terminology in this paper, these terms are used for
          no particular reason.  Cf.~footnote \ref{flarb footnote}.}
in no particular order.  Each such step has zero amortized cost.
Any number of such operations might need to be performed and we stop when
neither can be applied.  At this point, the anchored subtree $S'$
has a particular form and we perform a final operation,
called a ``stretch'', at the cost of $O(\log n)$ amortized pointer changes.
This bound, together with the observation that the potential drop over any
sequence of operations is $O(n \log n)$, gives the theorem.
We now describe the details of zig-zagging and stretching.

\paragraph{The zig.}

A \emph{zig} is executed whenever a light left edge is part of the anchored
subtree~$S'$; see Figure~\ref{fig:T1}.  The zig operation simply involves
a right rotation on the edge in question.  The actual cost of a zig is $O(1)$,
which we set to be $1$ to ease the analysis.

\begin{figure}
  \centering
  \scalebox{0.5}{\ifpdf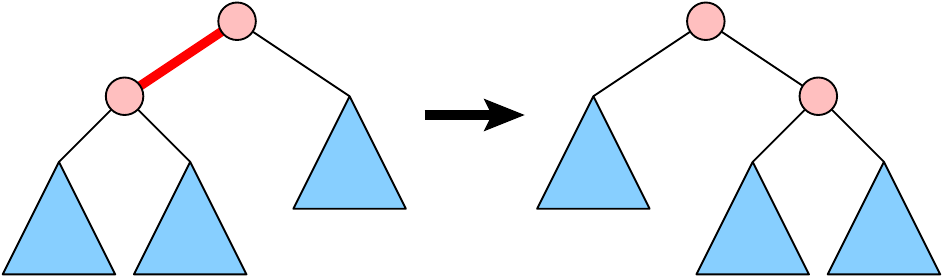\else\input{f1.pstex_t}\fi}
  \caption{A zig: The thick edge belongs to the anchored subtree~$S'$
           and is light.}
  \label{fig:T1}
\end{figure}

To analyze the change in potential, let $A$ and $B$ denote
the two children subtrees of the lower endpoint of the edge, and let $C$
denote the right child subtree of the upper endpoint of the edge.
In all formulas below, we use the same letters to denote the expanded weight
of the subtree.
Because the edge is light, $A+B+1 \leq C$.
Then the potential change is
\begin{align*}
\Delta \Phi & = 
\underbrace{
  \lg \frac{A}{B+C+1} + \lg \frac{B}{C}
}_{\text{new potential}}
~-~
\underbrace{
\left(
  \lg \frac{A}{B} + \lg \frac{A+B+1}{C} 
\right)
}_{\text{old potential}}
\\
& = \lg \underbrace{\frac{B}{A+B+1}}_{< 1} + 
\lg \underbrace{\frac{B}{B+C+1}}_{< \frac{1}{2}\text{ because }C>B}
< -1.
\end{align*}
Therefore, the amortized cost of a zig is (at most) zero, as claimed.

\paragraph{The zag.}

A \emph{zag} is performed whenever there exists,
within the anchored subtree~$S'$, a path that goes left one edge,
right zero or more edges, and then left again one edge;
see Figure~\ref{fig:T2}.
The zag operation performs a constant number of pointer changes to re-arrange
the path in question into a right-leaning path.
The actual cost of a zag is $O(1)$,
which we again set to be $1$ to ease the analysis.

\begin{figure}
  \centering
  \scalebox{0.5}{\ifpdf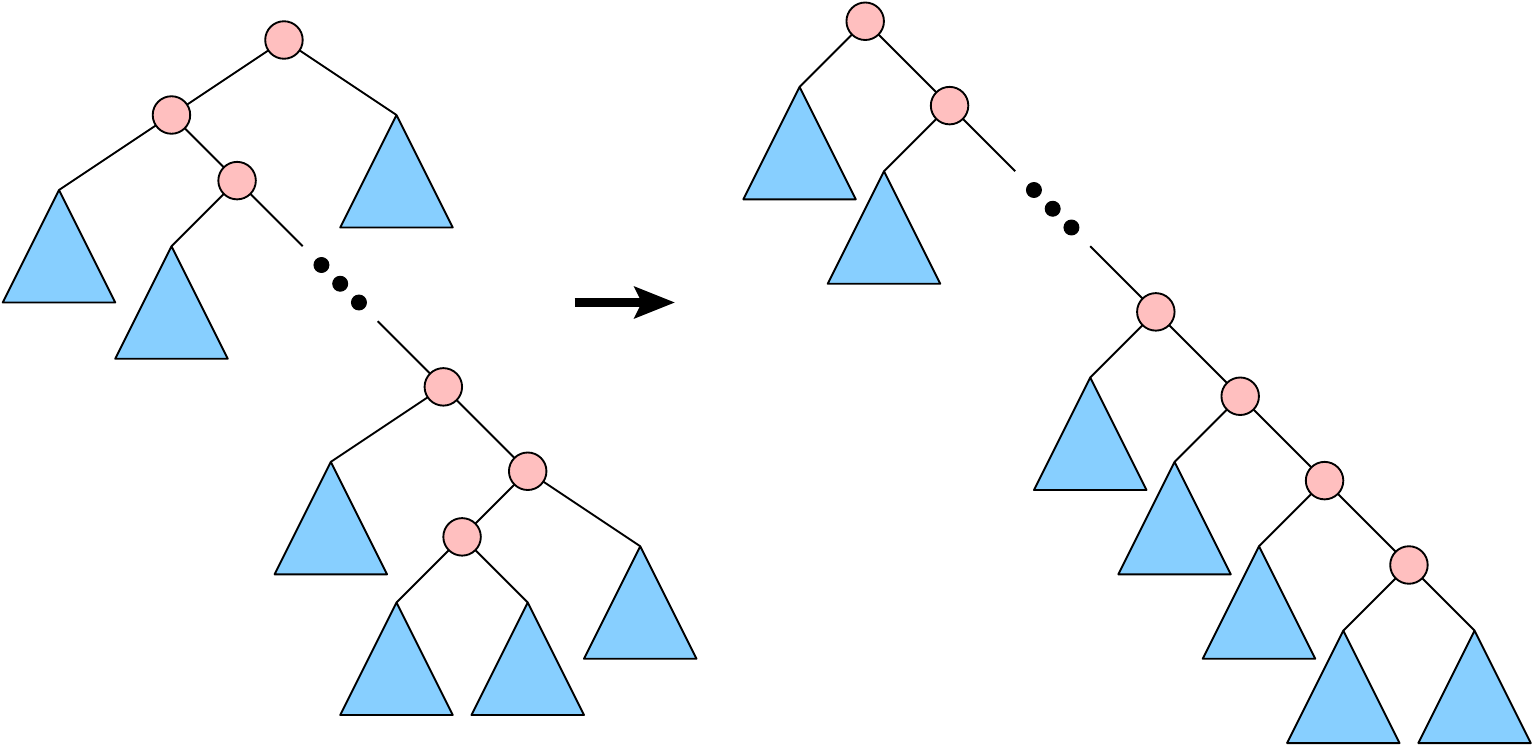\else\input{f3.pstex_t}\fi}
  \caption{A zag.}
  \label{fig:T2}
\end{figure}

We now argue that a zag reduces the potential by at least~$1$.
First, notice that the contribution to the potential of
parent nodes of the trees $B_1, B_2, \dots, B_k$ decreases after the
execution of the zag because, in each case,
the left subtree remains the same while the right subtree grows.
We will argue that the contribution of the remaining nodes decreases
by at least~$1$.  Indeed,
\begin{align*}
\Delta \Phi & \leq
\underbrace{
  \lg\frac{C}{D+E+F+2} + \lg\frac{D}{E+F+1} + \lg\frac{E}{F}
}_{\text{new potential}}
\\
& \quad {} - ~
\underbrace{
\left(
  \lg\frac{\sum_i B_i+C+D+E+k+2}{F} +  \lg\frac{C+D+1}{E} + \lg\frac{C}{D}
\right)
}_{\text{old potential}}
\\
& = \lg\frac{D^2 E^2}{\underbrace{(D+E+F+2)}_{> D+E}(E+F+1)\underbrace{(\Sigma_i B_i+C+D+E+k+2)}_{> D+E}(C+D+1)} \\
& < \lg\Big(\frac{1}{2} \cdot \underbrace{\frac{2D E}{(D+E)^2}}_{<1}
\cdot \underbrace{\frac{D}{C+D+1}}_{<1} \cdot \underbrace{\frac{E}{E+F+1}}_{<1}\Big)
< -1,
\end{align*}
as claimed.

\paragraph{The final stretch.}

After all possible zigs and zags have been exhausted,
we claim that the anchored subtree~$S'$
must have the form shown in Figure~\ref{fig:T3}.
\begin{figure}
  \centering
  \scalebox{0.5}{\ifpdf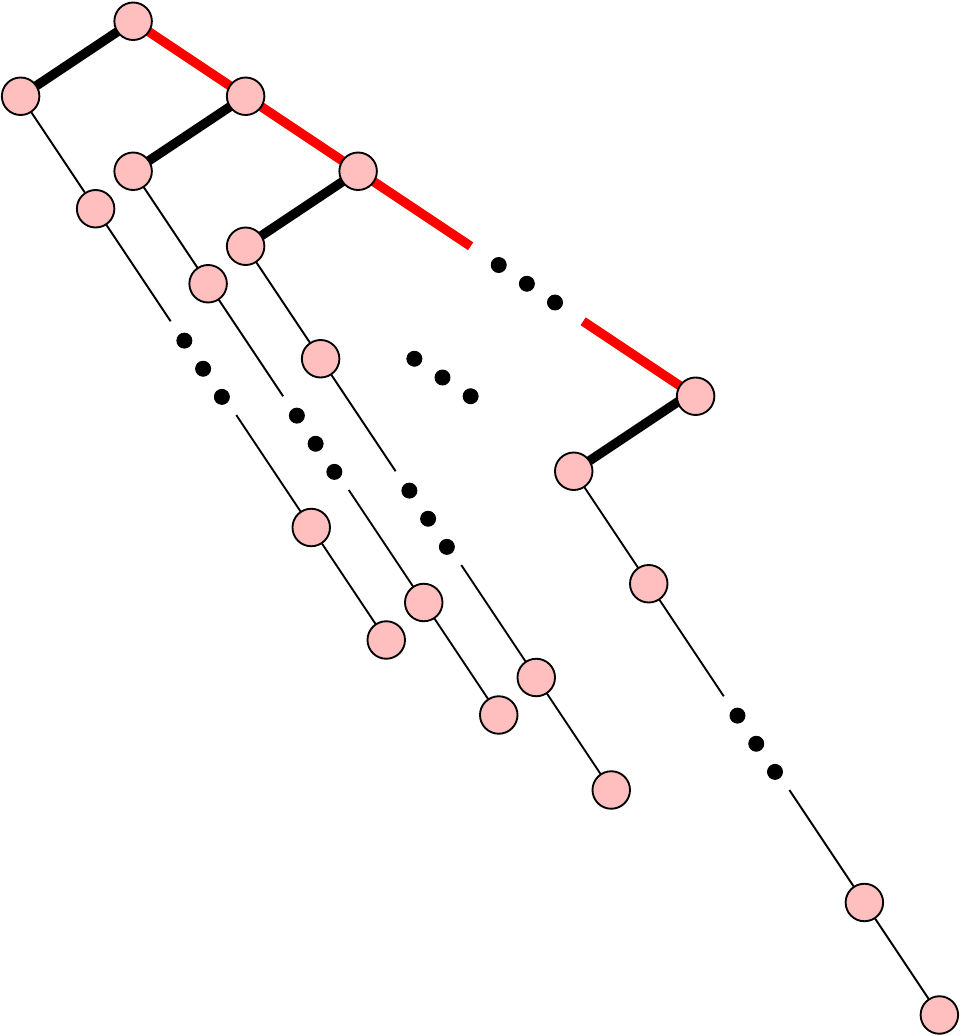\else\input{f4.pstex_t}\fi}
  \caption{The form of the anchored subtree $S'$ before the final stretch.
    The thick light edges are light, and the thick black edges are heavy.}
  \label{fig:T3}
\end{figure}
Indeed, any tree that has no light left edge and no right-leaning path
delimited by two left edges must have this form.
In particular, because the rightmost path in this tree must be light,
its length is at most $\lg (2 n + 1)$.

The final \emph{stretch} operation,
which completes the flarb, simply converts this tree into a rightmost path
by effectively concatenating the subsidiary right-leaning paths,
incorporating them into the main path.  Only $O(\log n)$ actual
pointer changes are required.  The potential does not increase because
left subtrees of every node shrink and right subtrees grow, if they
change at all.  Therefore, the amortized cost of the
stretch is indeed $O(\log n)$.

This concludes the proof of the theorem.
\end{proof}

\section{Transformations}

In this section we show how flarbs and grappa trees come together
with a little work to give us the main result.
In the next two transformations, we focus on the farthest-point case,
but the proofs apply equally well to the nearest-point version.

\begin{transform} \label{transform 1}
  Given a grappa tree data structure supporting each operation in
  $O(\log n)$ amortized time,
  and given a data structure to incrementally maintain a tree created by
  $n$ flarbs with $O(\log n)$ amortized pointer changes per flarb,
  we can construct an $O(n \log^2 n)$-space data structure that supports
  $O(\log n)$-time farthest-point queries on any prefix of a sequence of
  points in convex position in counterclockwise order.
\end{transform}

\begin{proof}
  We construct an incremental data structure that supports
  $O(\log n)$-time farthest-point queries on the current sequence of points,
  $\langle p_1, p_2, \dots, p_n \rangle$, and supports appending a new point
  $p_{n+1}$ to the sequence provided that this change maintains the invariant
  that the vertices remain in convex position and in counterclockwise order.
  Thus the insertion order equals the index order and equals the
  counterclockwise traversal order of a convex polygon.
  The data structure runs on a pointer machine
  in which each node has bounded in-degree.
  Thus we can apply the partial-persistence transform
  of \cite{Driscoll-Sarnak-Sleator-Tarjan-1989} and
  obtain the ability to support farthest-point queries
  on any prefix of the inserted points in $O(\log n)$ time.
  The space usage becomes proportional to the number of pointer changes
  during the insertions.

  We consider the expanded rooted binary tree $T$ formed by the edges of the
  farthest-point Voronoi diagram, ignoring their exact geometry;
  see Figure~\ref{farthest point}.
  To define $T$ precisely, recall that the \emph{farthest-point Voronoi diagram}
  \cite[Section~6.3]{Preparata-Shamos-1993}
  divides the plane into $n$ cells by classifying each point $q$ in the plane
  according to which of $p_1, p_2, \dots, p_n$ is the farthest from~$q$.
  The \emph{farthest-point Delaunay triangulation}
  \cite{Eppstein-1992-farthest}
  is the dual of the farthest-point Voronoi diagram, i.e., it triangulates
  the convex polygon with vertices $p_1, p_2, \dots, p_n$ by connecting
  two vertices whenever the corresponding Voronoi cells share an edge.
  (If a vertex of the Voronoi diagram has degree more than three, we
   conceptually split it into a tiny, arbitrarily chosen binary tree.)

  \begin{figure*}
    \centering
    \includegraphics[width=\linewidth]{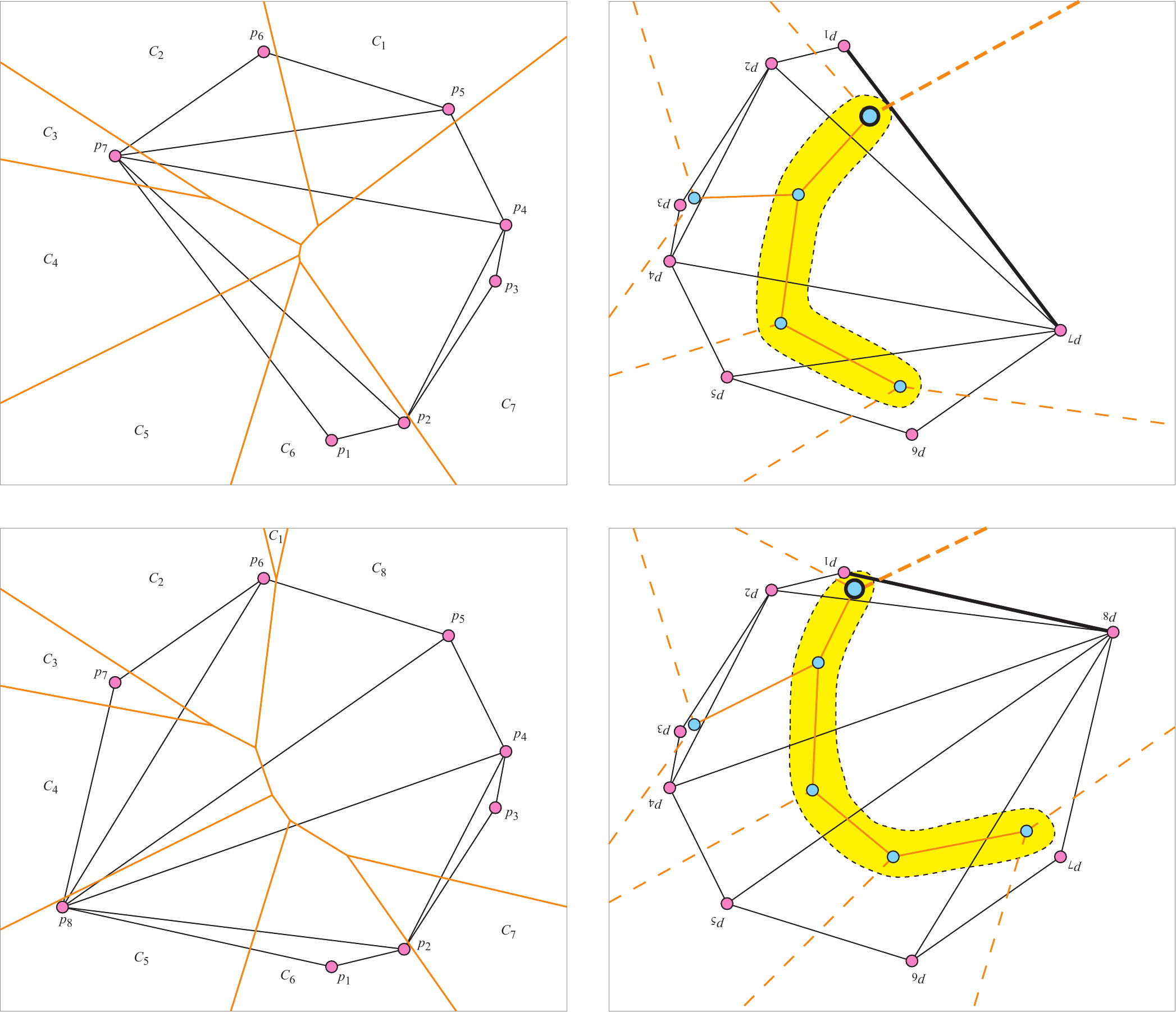}
    \caption{Adding point $p_8$ in counterclockwise order.
      Top: Before.  Bottom: After.
      Left: Farthest-point Voronoi diagram and its dual,
      the farthest-point Delaunay triangulation.
      Point $p_i$'s cell is denoted~$C_i$.
      Right: Delaunay triangulation and its dual, the tree~$T$,
      with infinite rays drawn as dashed lines,
      centrally reflected (rotated $180^\circ$)
      to roughly match the geometry on the left.
      Labels are also rotated $180^\circ$ for emphasis.
      The root vertex of $T$ and its parent edge are emboldened.}
    \label{farthest point}
  \end{figure*}

  Now define $T$ to be the expansion of the dual tree of this farthest-point
  Delaunay triangulation of the convex polygon (excluding the outside region),
  where each internal node in the tree represents a triangle in the
  farthest-point Delaunay triangulation, or equivalently,
  a vertex in the farthest-point Voronoi diagram.
  Each edge between internal nodes in $T$ corresponds to (a nongeometric
  representation of) a finite edge of the farthest-point Voronoi diagram,
  which bisects two of the points $p_i$ and $p_j$
  that are adjacent in the Delaunay triangulation.
  Each edge to an external node or superroot corresponds to
  an infinite ray of the farthest-point Voronoi diagram.
  Root the tree at the node corresponding to the unique triangle
  in the Delaunay triangulation bounded by the edge connecting the first
  inserted point $p_1$ and the most recently inserted point~$p_n$,
  so that the infinite ray emanating from this Voronoi vertex
  corresponds to the edge from this root node to the superroot.
  Define the notions of \emph{left child} versus \emph{right child} of a
  node according to the counterclockwise order around the Voronoi vertex.

  %
  %
  Define the \emph{left mark} of an edge
  to be the label of the region to the left of the edge, and symmetrically
  for the \emph{right mark}.
  Thus, the two marks of an edge define the two points $p_i$ and $p_j$
  whose bisector line contains the Voronoi edge.
  %
  The tree $T$ is not balanced, so we use a grappa tree to represent it
  and the left and right marks of edges.

  Next we consider the effect of inserting a new point~$p_{n+1}$.
  As in the standard incremental algorithm for Delaunay construction
  \cite[Section~9.3]{Berg-Kreveld-Overmars-Schwarzkopf-1999},
  we view the changes to the farthest-point Delaunay triangulation
  as first adding a triangle $p_1, p_n, p_{n+1}$ and then flipping
  a sequence of edges to restore the farthest-point Delaunay property.
  The key property of the edge-flipping process is that all flipped
  edges end up incident to the newly inserted point $p_{n+1}$.
  Therefore these changes can be interpreted in the tree as
  adding a new root node, whose left child is the previous root,
  and then choosing a collection of internal nodes to move to the right path
  of the new root.  This collection of nodes induces a connected subtree
  because the triangles involved in the flips form a connected set.
  (In particular, the flipping algorithm considers the neighbors of a triangle
  for flipping only if the triangle was already involved in a flip.)
  Thus, the changes correspond exactly to a flarb (in the unexpanded tree),
  with the flexibility of the flarb operation (choice of anchored subtree)
  encompassing the various possibilities of which
  edges get flipped to maintain the farthest-point Delaunay property.
  Another way to view the addition of $p_{n+1}$ is directly in the
  Voronoi diagram.  The point $p_{n+1}$ will capture the convex region
  $R_{n+1}$ for which $p_{n+1}$ is the farthest neighbor.
  Outside $R_{n+1}$, the Voronoi diagram
  is unchanged, so all edges of the new Voronoi diagram are either
  bisectors of the same two points as before, or are edges of~$R_{n+1}$.
  In $T$ after the flarb, $R_{n+1}$ corresponds to the right spine. 

  Each pointer change during a flarb operation can be implemented with
  one cut and one link operation.
  Therefore the grappa tree implements the $O(n \log n)$ total pointer updates
  from flarb operations in $O(n \log^2 n)$ total pointer updates.
  It remains to update the marks on the edges.
  By the incremental Voronoi/Delaunay view above, the only edges for which
  these marks might change are the edges incident to the new region~$R_{n+1}$,
  i.e., the edges on the right spine.
  We update the right marks on all of these edges by calling
  Right-Mark$(T,x,n+1)$ where $x$ is the rightmost vertex in~$T$,
  thus marking the entire right spine of~$T$.
  During the execution of the flarb, various right paths were cut and pasted
  together with cuts and links to form the final right spine.
  The edges on the final right spine that were originally part of a right path
  in $T$ already had their left mark set correctly.
  Any other edges on the final right spine were just added via links,
  so their left marks can be set by calling Left-Mark on the linked root
  just before calling Link.
  Thus, the total number of mark updates is also $O(n \log n)$,
  each costing $O(\log n)$ amortized.
  This concludes the space bound of the data structure.

  To support farthest-point queries, it suffices to build an oracle for
  the grappa tree's Oracle-Search.
  Specifically, given two incident edges $(u,v)$ and $(v,w)$, the oracle must
  determine which side of $(u,v)$ has the answer to the farthest-point query.
  Let $p_i$ and $p_j$ be the points defined by the two marks of the edge
  $(u,v)$; the two marks of edge $(v,w)$ define one of $p_i$ or $p_j$
  and a third point~$p_k$.
  Points $p_i$, $p_j$, and $p_k$ are the vertices of the
  Delaunay triangle corresponding to vertex $v$ in~$T$.
  The vertex of the Voronoi diagram corresponding to $v$
  can be computed as the intersection of the three perpendicular
  bisectors between these three points.
  We draw two rays from this Voronoi vertex in the direction opposite
  the two vertices $p_i$ and~$p_j$.
  (For a nearest-point Voronoi diagram, we draw rays perpendicular and toward
   supporting lines of the convex hull at $p_i$ and~$p_j$.)
  These two rays divide the plane into two sectors, and in constant time,
  we can decide which of the two sectors contains the query point~$q$.
  If the query point is in the sector containing the Voronoi edge
  corresponding to $(u,v)$,
  then the oracle returns the side of $T$ containing $(u,v)$, and vice versa.
  These rays, for every edge $(u,v)$ of $T$, subdivide the Voronoi cells into
  regions; the Oracle-Search will return the edge corresponding to the
  region containing the query point~$q$.
  In constant time, using the two labels on that edge of the tree,
  we can determine which side of the bisector contains~$q$, and therefore
  which farthest-point Voronoi region contains~$q$, i.e., which point $p_i$
  is farthest from~$q$.

  \xxx{Should analyze the construction time for the data structure.}

  This concludes the proof of the theorem.
\end{proof}

\begin{transform} \label{transform 2}
  Given an $O(n \log^2 n)$-space data structure that supports
  $O(\log n)$-time farthest-point queries
  on any prefix of a sequence of $n$ points ordered in convex position
  in counterclockwise order,
  we can construct an $O(n \log^3 n)$-space data structure that supports
  $O(\log n)$-time farthest-point-left-of-line queries on $n$ points
  in convex position.
\end{transform}


\begin{proof}
  Let $p_1, p_2, \dots, p_n$ denote the $n$ points in counterclockwise order.

  First we observe that, using the given prefix structure,
  we can also build an $O(n \log^2 n)$-space data structure that supports
  $O(\log n)$-time farthest-point queries on any suffix of a sequence of
  $n$ points ordered in convex position in counterclockwise order.
  We simply reflect the points about a fixed axis, reverse the order
  of the points, and build the prefix structure, and then apply the same
  reflection transformation to query points before giving it to the structure.

  Next we observe that, in $O(\log n)$ time, we can find the interval
  $p_i, p_{i+1}, \dots, p_j$ (where indices may wrap around modulo~$n$)
  of points that are to the left of the query line.
  This algorithm is described in the proof of Lemma~\ref{obs:basicAlg}.

  We build a collection of prefix and suffix data structures,
  and answer a query, via a divide-and-conquer recursion.
  The top level of the recursion is special because the sequence
  $p_1, p_2, \dots, p_n$ is cyclic.
  In this case we build a prefix structure and a suffix structure
  on this list of points.
  These structures can be used to solve any query interval that contains
  either $p_1$ or $p_n$ or both.
  Namely, if interval contains exactly one of $p_1$ or $p_n$,
  then the interval is a prefix or suffix of $p_1, p_2, \dots, p_n$.
  Otherwise, the interval is the union of a prefix and a suffix,
  so we can query both structures and return the farther of the two answers.

  At the general level of recursion, we have an interval
  $p_i, p_{i+1}, \dots, p_j$ of points, $i < j$, and a guarantee
  that any query interval reaching this level of recursion is strictly
  contained within this interval (excluding both $p_i$ and~$p_j$).
  At the top level of recursion, $i=1$ and $j=n$ and we know that
  the interval contains neither $p_1$ nor $p_n$ as required.
  Let $m = \lfloor (i+j)/2 \rfloor$ be the point midway between $i$ and~$j$.
  We construct a suffix data structure on the left half of points,
  $p_i, p_{i+1}, \dots, p_m$, and a prefix data structure on the right half
  of the points, $p_{m+1}, p_{m+2}, \dots, p_j$.
  As above, these data structures can be used to solve any query interval
  that contains either $p_m$ or $p_{m+1}$ or both
  (and satisfies the assumption of being strictly contained within the
  interval $p_i, p_{i+1}, \dots, p_j$).
  Then we recursively build data structures in the left half and in the
  right half for query intervals that do contain neither $p_m$ nor~$p_{m+1}$.
  In a query, we only need to recurse in one of the halves; we can decide
  which half overlaps the query interval in constant time by comparing $m$
  with the indices of the endpoints of the query interval.
  In the base case, $j=i$ or $j=i+1$ and there are no query intervals
  because of the strict containment, so there is nothing to do.

  The recurrence for query time is $T(n) = T(n/2) + O(1)$ plus an unknown
  base-case cost of $O(\log n)$, which solves to $O(\log n)$.
  The recurrence for space of the prefix and suffix data structures is
  $S(n) = 2  S(n/2) + O(n \log^2 n) = O(n \log^3 n)$.
  \xxx{Should analyze construction time here.  This is easy once we know
    the construction time for the previous transform.}
\end{proof}

Combining Theorems \ref{grappa theorem} and~\ref{flarb theorem}
with Transforms \ref{transform 1} and~\ref{transform 2},
we obtain the following main result of our paper:

\begin{corollary}
  There is an $O(n \log^3 n)$-space data structure that supports
  $O(\log n)$-time halfplane proximity queries on $n$ points
  in convex position.
\end{corollary}

We also mention the implication in the area of dynamic Voronoi diagrams,
which follows from combining Theorems \ref{grappa theorem}
and~\ref{flarb theorem} with Transform~\ref{transform 1}.

\begin{corollary} \label{Voronoi corollary}
  There is an $O(n)$-space data structure for maintaining a
  nearest-point or farthest-point Voronoi diagram of a sequence of points
  in convex position in counterclockwise order.
  The data structure supports inserting a new point at the end of the sequence,
  subject to preserving the invariants of convex position and counterclockwise
  order, in $O(\log n)$ amortized pointer changes per insertion;
  and supports point-location queries in $O(\log n)$ worst-case time.
\end{corollary}

\section{Open Problems and Conjectures}

Several intriguing problems remain open.
One obvious question is whether the $O(n \log^3 n)$ space of our
second data structure can be improved while keeping the optimal
$O(\log n)$ query time.  One specific conjecture in this direction
is the following:

\begin{conj}
  A sequence of $n$ flarb operations, starting from an empty tree,
  can be implemented at a cost of $O(1)$ amortized pointer changes per flarb.
\end{conj}
 
We have no reason to believe that our $O(\log n)$ amortized bound is tight.
Reducing the bound to $O(1)$ amortized would shave off a $O(\log n)$ factor
from our space and preprocessing time.
More importantly, it would increase our understanding of dynamic
Voronoi diagrams, reducing the $O(\log n)$ amortized update time
in Corollary~\ref{Voronoi corollary} to $O(1)$ amortized.
The potential function we use is inherently logarithmic;
a completely new idea is needed here for further progress. 

On the issue of improving our understanding of dynamic Voronoi diagrams,
we pose the following problem:

\begin{opp}
  Is there a data structure for maintaining a Voronoi diagram of a set of
  points in convex position that allows a point to be inserted in
  $\log^{O(1)} n$ time
  while supporting $O(\log n)$-time point location queries?
\end{opp}

Here we relax the condition that the points be inserted in
counterclockwise order, but maintain the restriction that
they be in convex position.  Although our potential
function does not give the result, it is possible that a slight variation
of it does.

Finally, it would be interesting to give explicit (and good) bounds on the
construction time in our second data structure, in particular so that it
completely subsumes the first data structure:

\begin{opp}
  Can the pointer changes caused by a flarb be found and implemented in
  $o(n)$ time, preferably $\log^{O(1)} n$ time?
\end{opp}

We have not been able to fully transform our combinatorial observations
about the number of pointer changes into an efficient algorithm,
because we lack efficient methods for finding which pointers change.
A solution to this problem would give us an explicit bound on the construction
time for our data structure,
and would provide a reasonably efficient dynamic Voronoi data structure for
inserting points in convex position in counterclockwise order.

\xxx{What about higher dimensions, as in \cite{bcdgs-eappt-02}?
Mention as open problem?}

\section*{Acknowledgments}

This work was initiated at the Schloss Dagstuhl Seminar 04091 on
Data Structures, organized by Susanne Albers, Robert Sedgewick, and
Dorothea Wagner, and held February 22--27, 2004 in Germany.
This work continued at the
Korean Workshop on Computational Geometry and Geometric Networks,
organized by Hee-Kap Ahn, Christian Knauer, Chan-Su Shin, Alexander Wolff,
and Ren\'e van Oostrum,
and held July 25--30, 2004 at Schloss Dagstuhl in Germany;
and at the 2nd Bertinoro Workshop on Algorithms and
Data Structures, organized by Andrew Goldberg and Giuseppe Italiano,
and held May 29--June 4, 2005 in Italy.
We thank the organizers and institutions hosting these workshops for
providing a productive research atmosphere.
We also thank Alexander Wolff for introducing the problem to us.

\let\realbibitem=\bibitem
\def\bibitem{\par \vspace{-1.2ex}\realbibitem}

\bibliography{paper}
\bibliographystyle{alpha}

\end{document}